\documentclass[
    aps,
    prb,
    reprint,
    superscriptaddress,
    showkeys,
    amsmath,
    amssymb,
    a4paper,
    10pt
]{revtex4-2}

\usepackage{float}
\usepackage{caption}
\usepackage{multirow}
\usepackage[
    colorlinks,
    linkcolor = black,
    anchorcolor = black,
    citecolor = black,
    filecolor = black,
    menucolor = black,
    runcolor = black,
    urlcolor = blue
]{hyperref}
\usepackage{subcaption}
\usepackage{graphicx}

\newcommand{\expdt}{\langle\delta t\rangle}
\newcommand{\exptau}{\langle\tau\rangle}
\newcommand{\expNC}{\langle N_C\rangle}
\newcommand{\expND}{\langle N_D\rangle}
\newcommand{\Fe}{Fe\textsubscript{55}}
\newcommand{\ns}{ns\textsuperscript{-1}}
\newcommand{\s}{s\textsuperscript{-1}}
\newcommand{\pow}[2]{#1\cdot 10^{#2}}

\begin{document}

\title{Dynamics of growing carbon nanotube interfaces\\probed by machine learning-enabled molecular simulations}

\author{Daniel Hedman}
    \email{daniel.hedman@ltu.se}
    \affiliation{Center for Multidimensional Carbon Materials (CMCM), Institute for Basic Science (IBS), Ulsan 44919, Republic of Korea} 

\author{Ben McLean}
    \email{ben.mclean2@rmit.edu.au}
    \affiliation{Center for Multidimensional Carbon Materials (CMCM), Institute for Basic Science (IBS), Ulsan 44919, Republic of Korea} 
    \affiliation{School of Engineering, RMIT University, Victoria 3001, Australia} 

\author{Christophe Bichara}
    \email{christophe.bichara@cnrs.fr}
    \affiliation{Centre Interdisciplinaire de Nanoscience de Marseille (CINaM), Centre National de la Recherche Scientifique (CNRS), Aix Marseille Université, 13288 Marseille, France} 

\author{Shigeo Maruyama}
    \email{maruyama@photon.t.u-tokyo.ac.jp}
    \affiliation{Department of Mechanical Engineering, The University of Tokyo, Tokyo 113-8656, Japan} 

\author{J. Andreas Larsson}
    \email{andreas.1.larsson@ltu.se}
    \affiliation{Applied Physics, Division of Materials Science, Department of Engineering Sciences and Mathematics, Luleå University of Technology, 971 87 Luleå, Sweden} 

\author{Feng Ding}
    \email{f.ding@siat.ac.cn}
    \affiliation{Center for Multidimensional Carbon Materials (CMCM), Institute for Basic Science (IBS), Ulsan 44919, Republic of Korea} 
    \affiliation{Department of Materials Science and Engineering, Ulsan National Institute of Science and Technology (UNIST), Ulsan 44919, Republic of Korea} 
    \affiliation{Faculty of Materials Science and Engineering/Institute of Technology for Carbon Neutrality, Shenzhen Institute of Advanced Technology, Chinese Academy of Sciences, Shenzhen 518055, China} 

\date{February 19, 2023}

\begin{abstract}
Carbon nanotubes (CNTs) are currently considered a successor to silicon in future nanoelectronic devices. To realize this, controlled growth of defect-free nanotubes is required. Until now, the understanding of atomic-scale CNT growth mechanisms provided by molecular dynamics simulations has been hampered by their short timescales. Here, we develop an efficient and accurate machine learning force field, DeepCNT-22, to simulate the complete growth of defect-free single-walled CNTs (SWCNTs) on iron catalysts at near-microsecond timescales. We provide atomic-level insight into the nucleation and growth processes of SWCNTs, including the evolution of the tube-catalyst interface and the mechanisms underlying defect formation and healing. Our simulations highlight the maximization of SWCNT-edge configurational entropy during growth and how defect-free CNTs can grow ultralong if carbon supply and temperature are carefully controlled.
\end{abstract}

\keywords{DeepCNT-22, machine learning force field, molecular dynamics, defect-free, single-walled carbon nanotube, growth, MLFF, DeePMD, DFT, VASP}

\maketitle

\section{Introduction}

Single-walled carbon nanotubes (SWCNTs) are hollow cylindrical structures composed of a single layer of sp\textsuperscript{2}-hybridized carbon atoms arranged in a hexagonal lattice, rolled up along a specific axis. Since their discovery over three decades ago\cite{iijima1991helical}, SWCNTs have garnered significant attention due to their potential to replace silicon as a semiconductor in transistors\cite{cao2015end-bonded, zhong2017gigahertz} and copper as a metallic conductor in electrical circuits\cite{behabtu2013strong}. To fully realize their technological potential, it is crucial to produce nanotubes that are free of defects. Irregularities in the hexagonal lattice that typically manifest as 5- and 7-membered rings, and negatively impact the electronic and optical properties\cite{charlier1996structural, charlier2002defects} and mechanical strength\cite{takakura2019strength} of SWCNTs.

The desire to produce ultralong, defect-free SWCNTs has driven extensive research efforts aimed at understanding the atomic-level mechanisms of growth. Although \textit{in situ} transmission electron microscopy has provided valuable insights\cite{stolojan2006controlled, hofmann2007situ, yoshida2008atomic-scale, wang2020precise, fan2021dynamic, ma2022nucleation}, a comprehensive atomic-level understanding of SWCNT growth has yet to be achieved through experimental measurements alone. To address this, computational studies, particularly molecular dynamics (MD) simulations, have been instrumental in uncovering the underlying mechanisms of SWCNT growth\cite{ding2004nucleation, ding2004molecular, amara2008understanding, ohta2009quantum, ribas2009nanotube, page2010qmmd, page2010mechanisms, page2015insights, khalilov2015atomic, xu2015atomistic, mclean2017catalytic, hisama2018growth, yoshikawa2019molecular, xu2021catalyst, mclean2021initial, mclean2022mechanism,qiu2022carbon}. From these it is known that gaseous carbon-based precursors decompose on a catalyst surface, releasing carbon atoms that form long chains which subsequently fold to form rings. These 5- and 6-memberd rings combine to form a graphitic structure, defined as the SWCNT-cap once a minimum of six 5-rings have formed\cite{xu2018kinetics}. A recent model, verified via density functional theory (DFT) calculations and MD simulations, explains why SWCNT-caps lift-off from the catalyst surface\cite{ding2022carbon}.

Historically, MD simulations have been methodologically limited in accurately exploring the timescales necessary for defect-free growth without additional bias. For instance, defect-free SWCNTs were observed during atomistic simulations using an empirical carbon-nickel force field\cite{xu2015atomistic}, but these simulations were assisted by a basin-hopping strategy to remove transition states and artificially heal any topological defects at the interface. Theoretical studies have instead focused on initial reactive pathways\cite{shibuta2013ab, khalilov2015atomic, mclean2021initial}, cap nucleation\cite{amara2008understanding, page2010qmmd, page2010mechanisms, neyts2011changing}, tube-catalyst interface\cite{larsson2007calculating, ding2007importance, silvearv2015establishing}, tube stability\cite{hedman2015stability, hedman2017length} and catalyst structure\cite{vets2017stabilities}. As a result, the growth of defect-free SWCNTs has not yet been achieved using pure MD simulations, and many questions related to growth remain unanswered, such as the timescale of the nucleation process, how defects form and heal, and the evolution of the tube-catalyst interface during growth, all of which are key to understanding growth of ultralong, defect-free carbon nanotubes.

An emerging and powerful method for modeling materials at experimentally relevant length and timescales is machine learning force fields (MLFFs)\cite{unke2021machine}. This technique involves training machine learning (ML) methods on a large dataset of atomic configurations that are labeled with energies, forces, and other properties calculated using highly accurate methods such as density functional theory (DFT). When trained, MLFFs can be used to predict material properties and drive atomistic simulations with the computational efficiency of empirical force fields, but with the accuracy of DFT\cite{qian2021comprehensive}. Recently, MLFFs have been developed to describe the bulk and nanostructure of various materials such as carbon\cite{rowe2020accurate}, silicon\cite{deringer2021origins}, phosphorus\cite{deringer2020general-purpose}, copper\cite{zhang2020dp-gen:} and gold\cite{li2022origin}, as well as more complex systems such as the phase diagram of water\cite{zhang2021phase}, liquid-liquid phase transitions\cite{yang2021liquid-liquid}, organic and inorganic reactions\cite{zeng2020complex, zeng2020exploring}, and high-entropy alloys\cite{xiang-guo2020complex}.

In this study, we present a novel MLFF, DeepCNT-22, which is suitable for MD simulations of SWCNT growth on iron catalysts. Using this method, we simulate the complete process of SWCNT growth at close to microsecond timescales without sacrificing computational accuracy and without the use of steering or other additional bias. Starting from a clean iron catalyst, we deposit carbon atoms one by one and achieve growth of nanometer-long, defect-free SWCNTs. Our simulations provide unprecedented atomic-level insights into the mechanisms of SWCNT growth, including the formation and healing of defects, which is key to growth of ultralong, defect-free SWCNTs. We also show, for the first time, the evolution of the tube-catalyst interface during growth and highlight the importance of the configurational entropy of the SWCNT-edge. To demonstrate the strength of this method, we report results from the growth of hundreds of SWCNTs which arise naturally from our MD simulations, without steering or other additional bias. These are used to study the chirality distribution of SWCNTs grown on iron catalysts, as well as the effect of catalyst size on the diameter of the grown tubes.

\section{Results and discussion}
\begin{figure*}
    \centering
    \includegraphics[width=\textwidth]{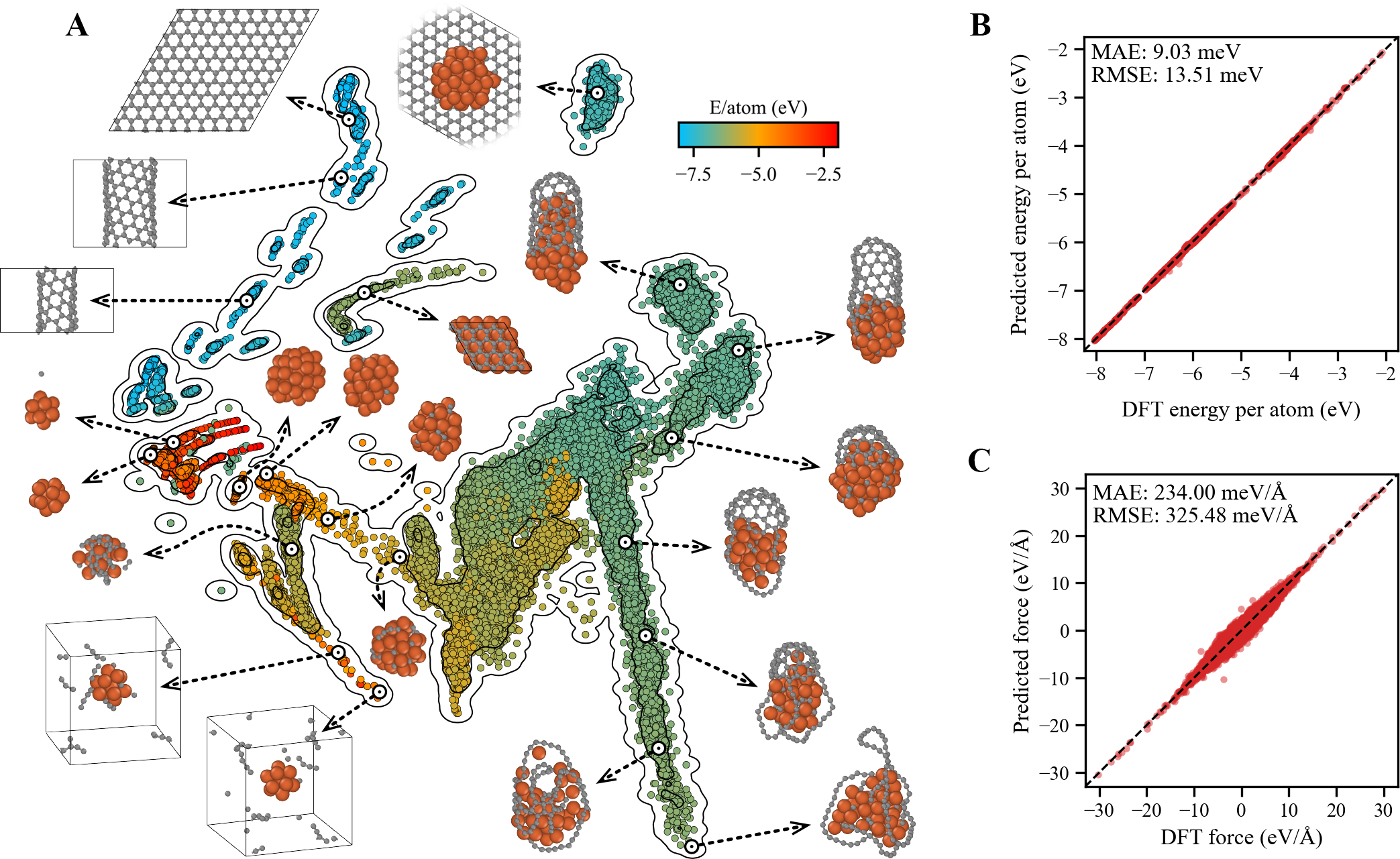}
    \caption{\label{fig:mlff} DeepCNT-22 dataset and accuracy verification. \textbf{A} sketch-map representation of the complete dataset (22 975 structures) used to train DeepCNT-22. Each colored dot represents a structure in the dataset, with the color corresponding to it's energy. The visualized atomic configurations show examples of structures from different regions of the sketch-map. \textbf{B} and \textbf{C} are the regression plots in energy and force, respectively, for DeepCNT-22 evaluated on the test data, which is a 10\% subset of the complete dataset shown in \textbf{A}. Here the atomic configurations are visualized using the OVITO software\cite{stukowski2009visualization}.}
\end{figure*}
\begin{figure*}
    \centering
    \includegraphics[width=\textwidth]{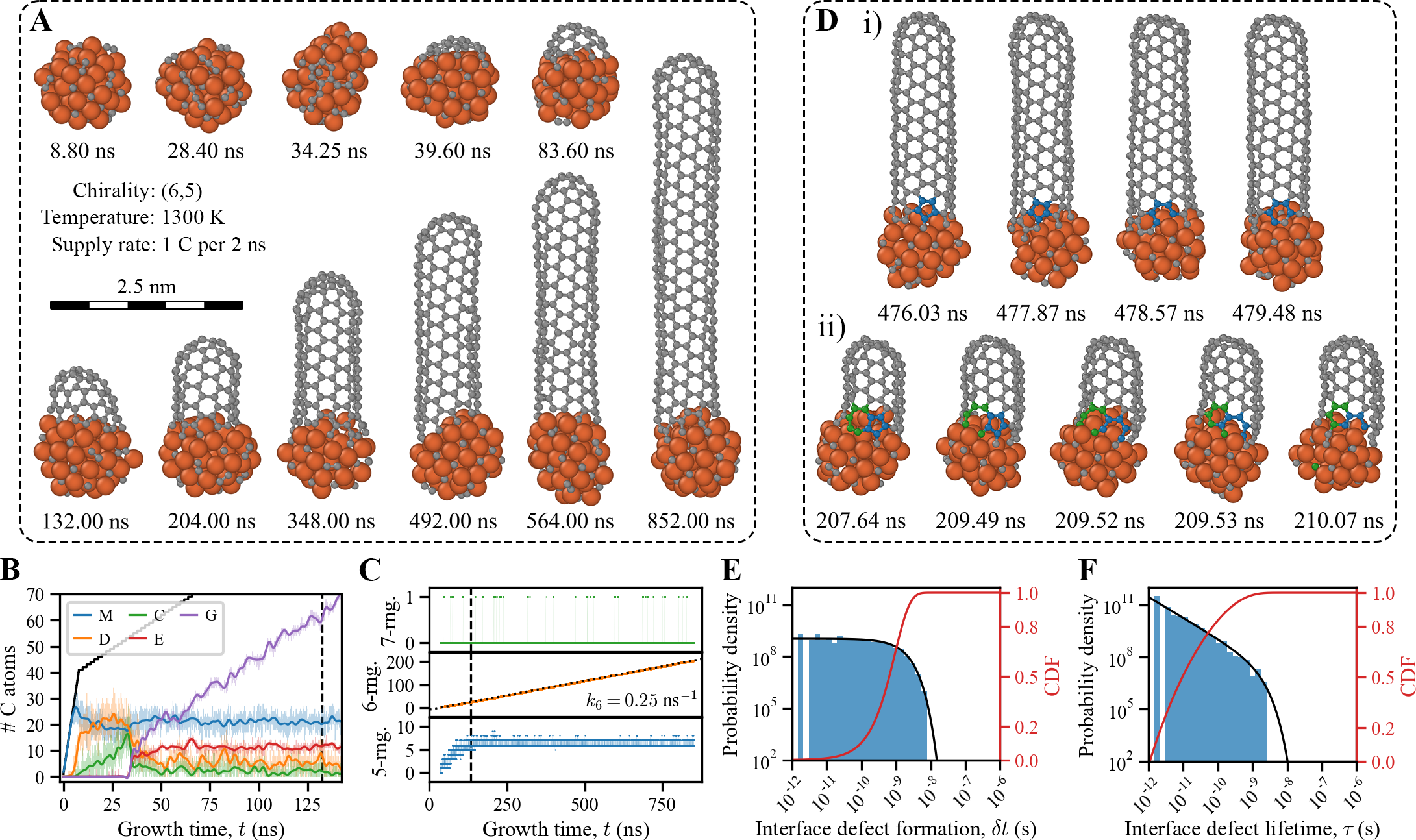}
    \caption{\label{fig:growth} Growth of a defect-free $(6,5)$ SWCNT on a \Fe{} catalyst at a temperature of $T=1300$ K and a growth rate of $k=0.5$ \ns{}. Panel \textbf{A} displays 11 snapshots of the structure during the growth process, and panel \textbf{D} illustrates the healing of interface defects i) 5-ring and ii) 5-7 pair. The orange and gray spheres represent Fe and C atoms, respectively, in panel \textbf{D}, blue and green spheres depict C atoms initially belonging to 5-ring and 7-ring interface defects, respectively. \textbf{B} shows the number of carbon atoms comprising each species, including monomers (M), dimers (D), chains (C), part of the edges (E), and graphitic structures (G), during the early stages of growth. The solid black line is the total number of carbon atoms added to the system, the transparent colored lines represent raw data, and the solid lines is the result of applying a low-pass filter. \textbf{C} presents the number of 5-, 6-, and 7-rings during growth, with a linear regression (dotted line) determining the 6-ring formation rate, $k_6=0.25$ \ns{}. The dashed vertical line in \textbf{B} and \textbf{C} marks the time at which the SWCNT-cap is fully formed, $t=132.41$ ns. \textbf{E} and \textbf{F} show the probability density function (solid black line) and the cumulative distribution function (CDF) (solid red line) for the time between formation of interface defects, $\delta t$, and the interface defect lifetime, $\tau$, during the growth process after the cap is fully formed.}
\end{figure*}

Carbon nanotubes (CNTs) exhibit a low growth rate, equivalent to approximately one carbon atom being added to the growing tube every microsecond depending on it's diameter. Modeling the complete growth of reasonably long CNTs necessitates simulations on very long timescales, thereby demanding a computationally efficient (fast) many-body interatomic potential. To meet this requirement, we developed a MLFF (DeepCNT-22) using the Deep Potential Smooth Edition method\cite{zhang2018deep, zhang2018end-to-end}, as implemented in DeePMD-kit\cite{wang2018deepmd-kit:}. This approach has been proven to be both accurate and computationally efficient\cite{jia2020pushing}, with the potential for further acceleration through the utilization of non-von Neumann architectures\cite{mo2022accurate}.

A key challenge in developing MLFFs is the generation of high-quality, diverse datasets for training. Such datasets consist of atomic configurations (structures) labeled with energy, force and virial values obtained using highly accurate methods such as DFT. Our initial dataset, which includes multiple structures relevant to the growth of SWCNTs, was generated using a variety of methods, including MD driven by density functional tight binding\cite{elstner1998self-consistent-charge}, randomly perturbed structures, and carbon allotropes from the GAP-20 dataset\cite{rowe2020accurate, csanyi2020carbon}. Regardless of the generation method, all structures were labeled with energies and forces obtained via high-accuracy, dispersion-corrected\cite{klimes2011van} DFT calculations using the Vienna Ab initio Simulation Package\cite{kresse1993ab, kresse1996efficiency, kresse1996efficient}. The initial dataset was then refined using a variant of the active learning scheme\cite{podryabinkin2017active, smith2018more, zhang2019active} in which an ensemble of 5 MLFFs is trained on the initial dataset and used to drive MD simulations of SWCNT growth. During these simulations, the deviation in the MLFFs' predictions (i.e., model deviation) is used to identify unrepresented structures that appear during the growth process. These are then labeled and added to the initial dataset, and a new ensemble of MLFFs is trained. This process is repeated until the model deviation stays low throughout the SWCNT growth simulation. For more details on data labeling and the architecture/training of DeepCNT-22 we refer to the Methods section.

The dataset used to train DeepCNT-22, consisting of 22 975 structures, is depicted in Fig.~\ref{fig:mlff}A as a sketch-map representation\cite{ceriotti2011simplifying, fraux2020chemiscope:}. Each point in the map denotes a structure in the dataset, the position of which is determined by principal component analysis on the sum of the learned descriptors of the local atomic environments (embeddings) of all atoms in the structure. Different regions in the sketch-map correspond to different types of structures, as shown by the visualized atomic configurations. These regions show clear grouping of similar structures and clear separation of dissimilar  structures, indicating good diversity in the dataset and that relevant descriptors are learned by DeepCNT-22. 10\% of the dataset was withheld from the training process and used to test the accuracy of DeepCNT-22. As shown in the regression plots of Fig.~\ref{fig:mlff}B,C, DeepCNT-22 can reproduce the DFT energies and forces of the test dataset with an RMSE of 13.51 meV per atom for energies and 325.48 meV/Å for forces, similar to other published MLFFs based on the same architecture\cite{zeng2020complex, zeng2020exploring}. Further verification of DeepCNT-22 can be found in Fig. S1 of the Supplementary material.

Having verified the accuracy of DeepCNT-22, we used it in combination with the Large-scale Atomic/Molecular Massively Parallel Simulator\cite{thompson2021lammps} (LAMMPS) to simulate the growth of SWCNTs. The combination of DeepCNT-22 and LAMMPS enables us to simulate growth on \Fe{} catalysts at a rate of $\sim$70 ns/day using a single Nvidia V100 GPU and $\sim$1200 ns/day using a non-von Neumann architecture\cite{mo2022accurate}. While the non-von Neumann architecture achieves a speedup of $\sim$17 times, all simulations were ultimately performed on Nvidia K80 and V100 GPUs due to their greater availability. Nevertheless, the use of non-von Neumann architectures highlights the potential for MLFFs to simulate materials at ultralong timescales.

Growth simulations were performed as follows, starting from a pure \Fe{} catalyst, 40 carbon monomers were added at an initial supply rate of 5 C per ns, $k=5$ \ns{}, to saturate the catalyst. After which the carbon supply rate was reduced to between $k=0.5$ to $1.0$ \ns{} to reproduce the slow growth of SWCNTs. We found that this growth rate is low enough to reliably produce defect-free tubes while allowing for growth of long ($\sim$500 atoms) SWCNTs in a reasonable time ($\sim$14 to 7 days). To precisely control the supply of carbon atoms, we bypass the decomposition of the feedstock gas and directly add carbon monomers to the system. Due to the high solubility of carbon in iron and to prevent highly reactive monomers from colliding with the tube wall, leading to unwanted defects, we add carbon monomers to the center of the catalyst as has been done previously\cite{ding2004nucleation, ding2004molecular}. The low supply rate allows ample time for the deposited carbon atom to diffuse away from the center of the catalyst before another carbon atom is added. In fact, the supply rate is low enough that a 1:1 correlation between the supply rate and the growth rate is achieved; therefore, we use the terms supply rate and growth rate interchangeably.

To determine an ideal temperature range for growth, we performed a series of simulations at $k=1.0$ \ns{} and different growth temperatures, $T$. The results, detailed in Fig. S2, show that temperatures between 1200 K and 1500 K results in well-defined SWCNTs. Lower temperatures tend to produce highly defective tubes, while higher temperatures result in more encapsulated catalysts. Growth temperature and carbon supply rate are, thus, intertwined parameters that can be systematically investigated on realistic timescales using DeepCNT-22.

Growing SWCNTs on a \Fe{} catalyst using the procedure described above, with $k=0.5$ or 1.0 \ns{} and $T=1300$ or 1500 K, we were able to produce several long, defect-free tubes. It should be noted that the chirality of these tubes was not predetermined, but rather emerged from the growth simulations. Fig.~\ref{fig:growth} shows the result of a 4.88 nm long $(6,5)$ SWCNT grown over 0.852 \textmu s at $T=1300$ K and $k=0.5$ \ns{}. This corresponds to a growth rate of 5730 \textmu m/s, which is approximately 50 to 1000 times higher than experimentally reported growth rates\cite{yao2007raman, unrau2009single-walled, huang2011review, jourdain2013current} and around 10 to 100 times lower than previous MD studies\cite{ding2004nucleation, ribas2009nanotube, xu2015atomistic, hisama2018growth}. Despite the high growth rate, the resulting SWCNT shown in Fig.~\ref{fig:growth}A is free of defects, demonstrating that growth of defect-free tubes can be achieved even at very high growth rates compared to experiments. This is due to the matching of the carbon supply rate with the growth temperature, as will be detailed later. Additional examples of long, defect-free SWCNTs resulting from our growth simulations can be found in Fig. S3.

In the following sections, we will investigate the mechanisms behind the growth of defect-free SWCNTs using the $(6,5)$ tube as a case study.

\subsection{The five stages of nanotube growth}

The growth process of the $(6,5)$ SWCNT can be observed in Fig.~\ref{fig:growth}A and \href{https://youtu.be/K90Ca6uDNEQ}{Supporting video S1} from which we see the following. During the earliest (1\textsuperscript{st}) stage of growth, the dominating carbon species are monomers and dimers. This is represented by the snapshot at $t=8.80$ ns and can be seen by the carbon species analysis shown in Fig.~\ref{fig:growth}B from which we conclude that there are almost no carbon chains present on the catalyst during the 1\textsuperscript{st} stage. It is worth noting that the high carbon supply rate used at the beginning of growth, $k=5$ \ns{}, results in the 1\textsuperscript{st} stage lasting significantly less time than if a slower rate of $k=0.5$ \ns{} had been used.

We characterize the 2\textsuperscript{nd} stage of growth by the conversion of carbon monomers and dimers into linear carbon chains as seen in the snapshot structure at $t=28.4$ ns and the steady increase in the number of carbon atoms as part of chains, as seen in Fig.~\ref{fig:growth}B. Once the number of carbon atoms as part of chains has reached a critical ratio of $\sim$1/3, the growth enters the 3\textsuperscript{rd} stage. Here, a junction is formed on a long chain by either a monomer, dimer or the end of another chain attaching and forming a three-coordinated carbon atom. This causes the resulting chain-like structure to fold in on itself and form the first carbon ring seen in the snapshot at $t=34.25$ ns. Formation of the first carbon ring promotes subsequent ring formation at a rapid rate as evident by the sharp decrease in the number of atoms in chains and the simultaneous sharp increase in the number of graphitic carbons as seen in Fig.~\ref{fig:growth}B. This results in the almost complete elimination of linear carbon chains (Fig.~\ref{fig:growth}B) as they become part of the growing graphitic structure seen at $t=39.60$ ns.

The growth then enters the 4\textsuperscript{th} stage, where the graphitic structure grows larger through the attachment of carbon monomers and dimers to it's edge. As the graphitic structure expands, it's curvature increases and, since it is energetically favorable to have a near-perpendicular interface between the catalyst and the graphitic structure\cite{ding2022carbon}, it lifts off the catalyst. The SWCNT-cap is fully defined once at least six 5-rings have formed inside the graphitic structure\cite{xu2018kinetics}, this occurs at $t=132.41$ ns, after which the growth enters the 5\textsuperscript{th} and final stage, which is the continuous elongation of the tube by the attachment of carbon atoms at it's edge.

Taken together, these five steps, combined with the snapshots in Fig.~\ref{fig:growth}A and \href{https://youtu.be/K90Ca6uDNEQ}{Supporting video S1}, provide a comprehensive atomic-level understanding of SWCNT nucleation and growth.

\subsection{Defect formation and healing during growth}
\begin{table*}
    \centering
    \caption{\label{tab:defect_stat} Defect statistics obtained from growth simulations of a $(6,5)$ SWCNT on a \Fe{} catalyst at different conditions. Here Sim. represents the different simulations, $T$ the growth temperature, $k$ the carbon supply rate and $t_\mathrm{end}$ the growth time. \# 5-rng., \# 7-rng. are the number of unique 5-rings and 7-rings formed during growth, respectively. $\expdt$ and $\exptau$ are the expectation values for the time between interface defect formation and interface defect lifetime, respectively. $F_{\delta t}^{-1}(0.99)$ and $F_{\tau}^{-1}(0.99)$ are the quantile functions (inverse CDFs), evaluated at a probability of 99\%, for the time between interface defect formation and interface defect lifetime, respectively.}
    \begin{tabular}{c c c c c c c c c c}
        \multirow{2}{*}{\textbf{Sim.}} & $T$ & $k$ & $t_{end}$ & \multirow{2}{*}{\textbf{\# 5-rng.}} & \multirow{2}{*}{\textbf{\# 7-rng.}} & $\expdt$ & $\exptau$ & $F_{\delta t}^{-1}(0.99)$ & $F_{\tau}^{-1}(0.99)$\\
        & \textbf{(K)} & \textbf{(ns\textsuperscript{-1})} & \textbf{(ns)} &  &  & \textbf{(ns)} & \textbf{(ns)} & \textbf{(ns)} & \textbf{(ns)}\\
        \hline\hline
        1 & 1300 &               0.5 &  852 &  778 &  26 & 0.925 & 0.082 &  4.26 & 1.17\\
        2 & 1500 &        $<10^{-3}$ & 1000 & 4648 & 273 & 0.215 & 0.028 &  0.99 & 0.43\\
        3 & 1400 &        $<10^{-3}$ & 1000 & 2485 &  89 & 0.402 & 0.036 &  1.85 & 0.59\\
        4 & 1300 & $<8\cdot 10^{-4}$ & 1283 & 1287 &  41 & 0.996 & 0.045 &  4.59 & 0.80\\
        5 & 1200 &        $<10^{-3}$ & 1000 &  413 &   3 & 2.415 & 0.053 & 11.12 & 0.95\\
        6 & 1100 & $<5\cdot 10^{-4}$ & 2000 &  237 &   4 & 8.382 & 0.061 & 38.60 & 1.05
    \end{tabular}
\end{table*}

As seen in the snapshot at $t=852.00$ ns in Fig.~\ref{fig:growth}A the final grown SWCNT is straight and of single chirality. This is only possible if there are no carbon rings other than 6-rings inside the tube wall. However, we will show that this is not because only 6-rings are formed at the 5\textsuperscript{th} stage of growth.

By analyzing the number of 5-, 6- and 7-rings during the 5\textsuperscript{th} stage of growth (to the right of the dashed line in Fig.~\ref{fig:growth}C), we see the continuous increase in the number of 6-rings and interestingly the rate of formation of 6-rings, $k_6$, is half of the growth rate i.e., $k_6=k/2=0.25$ \ns{} as shown in Fig.~\ref{fig:growth}C. Furthermore, the number of 5-rings frequently exceeds the six 5-rings that are part of the SWCNT-cap. Likewise the number of 7-rings is at times greater than zero, which means that both 5-, 6- and 7-rings are formed during the 5\textsuperscript{th} stage of growth. By analyzing the structure during growth (Supporting videos \href{https://youtu.be/K90Ca6uDNEQ}{S1}, \href{https://youtu.be/x8Z5Go5iW58}{S2} and \href{https://youtu.be/e1Yx14PQjkg}{S3}), we find that like the 6-rings, the 5- and 7-rings are formed at the interface between the tube and the catalyst. So, we distinguish here between interface defects (5- and 7-rings near the tube-catalyst interface) and trapped defects (5- and 7-rings trapped inside the tube wall).

Having shown that we achieve growth of defect-free SWCNTs (Fig.~\ref{fig:growth}A and S3) and confirmed the existence of both 5- and 7-rings during the 5\textsuperscript{th} stage of growth (Fig.~\ref{fig:growth}C). We conclude that these interface defects must be effectively healed during the growth process so to not get trapped within the tube wall.

Fig.~\ref{fig:growth}D i) shows an example of how a 5-ring interface defect was healed during the growth of the $(6,5)$ SWCNT. Here a 5-ring (shown in blue) forms at $t=476.03$ ns and is partially incorporated into the tube wall at $t=477.87$ ns by the formation of a complete 6-ring at the bottom left side of the interface defect. At this point the 5-ring still has one of it's sides exposed to the catalyst. The addition of one more carbon atom at the edge (to the bottom right of the interface defect) is required for the 5-ring to be fully incorporated within the tube wall. Before this has a chance to occur, at $t=478.57$, a metal atom facilitates carbon-carbon bond cleavage at the side of the 5-ring which is exposed to the catalyst. Now the interface defect is considered healed as the metal atom holds the ring open until a carbon atom replaces it, and a complete 6-ring is formed at $t=479.48$ ns. The entire process of defect formation and healing can be viewed in \href{https://youtu.be/x8Z5Go5iW58}{Supporting video S2}.

Another example of how an interface defect can be healed is shown in Fig.~\ref{fig:growth}D ii). Here a more complex 5-7 interface defect is formed during the growth of the $(6,5)$ SWCNT. Although this interface defect is more complex and requires more steps to heal compared to a 5-ring interface defect, the basic mechanisms are the same. The 5-7 interface defect is formed at $t=207.64$ ns and at $t=209.49$ ns the 7-ring is opened with the help of a metal atom. Shortly thereafter at $t=209.52$ ns the 5-ring is opened by the same mechanism and the left side of the open 7-ring attaches to the left side of the open 5-ring forming a 6-ring as seen at $t=209.53$ ns. The carbon atoms attached to the open 5-ring are then dissolved back into the catalyst and the remaining carbon atoms, which were part of the 5-ring, forms an armchair edge as seen at $t=210.07$ ns. The entire process of defect formation and healing can be viewed in \href{https://youtu.be/e1Yx14PQjkg}{Supporting video S3}.

By identifying and labeling each unique carbon ring that forms, we can track the formation and lifetime of interface defects during growth. Through MD using DeepMD-22, we can then, for the first time, perform growth simulations long enough to collect data for a statistical analysis of interface defect formation during nanotube growth (note that no trapped defects are observed in this case).

During the growth of the $(6,5)$ SWCNT shown in Fig.~\ref{fig:growth}A, a total of 778 unique 5-rings were formed, but only 26 unique 7-rings. For the unique 5-rings the measured time between formation of interface defects, $\delta t$, is plotted in Fig.~\ref{fig:growth}E as a log-log histogram. Here it is clear that $\delta t$ can be modeled using a typical exponential distribution which has a probability density function (PDF) of $f_{\delta t} = \lambda_1 e^{-\lambda_1\delta t}$. This is reasonable since the exponential distribution describes the time between events in a Poisson point process and thus the formation of interface defects can be considered as a purely stochastic process. Fitting the cumulative distribution function (CDF), $F_{\delta t} = 1 - e^{-\lambda_1\delta t}$, to the normalized cumulative sum of the measured values of $\delta t$ gives $\lambda_1 = \pow{1.08}{9}$ \s{} and an expected value for the interval between interface defect formation of $\expdt = 1/\lambda_1 = 0.925$ ns. From the CDF it is also evident that there is a 99\% probability that interface defect form within 4.26 ns of each other i.e., $F_{\delta t}^{-1}(0.99) = 4.26$ ns for the growth of a $(6,5)$ SWCNT at 1300 K.

Similarly, we can measure the lifetime of interface defects, $\tau$, which is shown in Fig.~\ref{fig:growth}F. Here the measured values follow a straight line in the log-log histogram which is the signature of a power law distribution i.e., $f_\tau\propto\tau^{-\alpha}$. This distribution is known to be heavy tailed meaning that the tail of the power law distribution is not exponentially bound\cite{asmussen2008steady-state}. However, as seen in Fig.~\ref{fig:growth}F, this is not the case for $\tau$ as there are no interface defects that have a lifetime longer than around 5 ns. Thus we model the interface defect lifetimes as a power law distribution with an exponential cutoff whose PDF is given as, $f_\tau = \frac{\lambda_2^{1-\alpha}}{\Gamma(1-\alpha,\:\lambda_2\tau_\mathrm{min})}\tau^{-\alpha}e^{-\lambda_2\tau}$, where $\Gamma(1-\alpha,\:\lambda_2\tau_\mathrm{min})$ is the upper incomplete gamma function, for details on the derivation of this PDF see the Supplementary material. Fitting the CDF, $F_\tau = 1 - \frac{\Gamma(1-\alpha,\:\lambda_2\tau)}{\Gamma(1-\alpha,\:\lambda_2\tau_\mathrm{min})}$, of this distribution to the normalized cumulative sum of the measured values of $\tau$ gives $\alpha = 1.20$, $\lambda_2 = \pow{1.04}{9}$ \s{} and $\tau_\mathrm{min} = \pow{1.10}{-12}$ s. With these we can determine the expected value of the interface defect lifetime as $\exptau = \frac{1}{\lambda_2}\frac{\Gamma(2-\alpha,\:\lambda_2\tau_\mathrm{min})}{\Gamma(1-\alpha,\:\lambda_2\tau_\mathrm{min})} = 0.082$ ns and from the CDF we see that 99\% of all interface defects have a lifetime shorter than 1.17 ns, $F_{\tau}^{-1}(0.99) = 1.17$ ns.

With these distributions we can model interface defects and predict both the time between formation, $\delta t$, and their lifetimes, $\tau$. To explore how $\delta t$ and $\tau$ are affected by different growth conditions such as carbon supply rate (growth rate), $k$, or temperature, $T$, we extracted a snapshot from the growth simulation of the $(6,5)$ SWCNT and performed MD simulations for 1 to 2 \textmu s at different temperatures without adding any carbon atoms to the system, Sim. 2-6 in Table~\ref{tab:defect_stat}. These simulations mimic conditions closer to that of experimental growth where the growth rate is around 50 to 1000 times lower than in our growth simulations.

The high growth rate of the $(6,5)$ SWCNT shown in Fig.~\ref{fig:growth} (Sim. 1 in Table~\ref{tab:defect_stat}) results in a $\sim$7\% reduction in both $\expdt$ and $F_{\delta t}^{-1}(0.99)$ compared to the MD simulation of the extracted snapshot at the same temperature (Sim. 4 in Table~\ref{tab:defect_stat}). Sim. 1 also show longer interface defect lifetimes, $\sim$82\% larger $\exptau$ and $\sim$46\% larger $F_{\tau}^{-1}(0.99)$ compared to Sim. 4. However, given that the growth rate of Sim. 1 is $>640$ times greater than that of Sim. 4 we conclude that both the time between formation of interface defects and their lifetimes are largely independent of the growth rate of the nanotube.

Comparing the results from the MD simulation of the extracted snapshot done at different temperatures, Sim. 2-6 in Table~\ref{tab:defect_stat}, we see that as the temperature decreases, the time between formation of interface defects increases significantly, with a 2 to 3 times increase in both $\expdt$ and $F_{\delta t}^{-1}(0.99)$ for a 100 K decrease in growth temperature. Similarly, both $\exptau$ and $F_{\tau}^{-1}(0.99)$ increase with a decrease of the temperature, although here the effect is less pronounced with only a 15 to 30\% increase for a 100 K decrease in temperature.

Given that the formation of interface defects is a stochastic process ($\delta t$ follows an exponential distribution), it can be modeled using the Arrhenius equation. By fitting this equation to $k_5=1/\expdt$ for Sim. 2-6, we obtain an activation energy of 1.30 eV for the formation of interface defects. On the other hand, the lifetime of interface defects is not a simple stochastic process, as evidenced by the fact that $\tau$ follows a power law distribution. This suggests the presence of a hierarchical structure, and as shown in Fig.~\ref{fig:growth}D, the healing of an interface defect is a complex process involving multiple steps. Therefore, the healing of interface defects cannot be accurately modeled using a simple Arrhenius equation.

\subsection{Expected length of defect-free CNTs}

Although it has been shown experimentally that centimeter-long CNTs can be grown without a single defect in the tube wall\cite{wen2010growing, wen2010100mm, zhang2011superstrong, zhang2013growth}, how growth of such ultralong defect-free CNTs is possible remains unclear. However, with our newfound insights into the formation and healing of interface defects during growth, we can reveal the key to growth of ultralong defect-free CNTs.

We propose a simple model to determine the expected length of a CNT in terms of the number of carbon atoms, $\expNC$, that it can reach during growth before an interface defect is likely to get trapped in the tube wall. Let us assume that $n_6$ 6-rings need to form at the interface of the growing tube to trap an interface defect inside the tube wall. For a given growth rate, $k$, the average formation rate of 6-rings during growth is equal to $k_6=k/2$ as shown in Fig.~\ref{fig:growth}C. Thus, the critical time that an interface defect must live to become trapped inside the tube wall is
\begin{equation}
    \tau_c = \frac{n_6}{k_6} = \frac{2n_6}{k}.
\end{equation}

Since we know the CDF, $F_\tau$, for the interface defect lifetimes, we can calculate the probability, $p$, that an interface defect lives for at least time $\tau_c$ using the survival function
\begin{equation}
    p = S_\tau(\tau_c) = 1 - F_\tau(\tau_c) = \frac{\Gamma(1-\alpha,\:\lambda_2\tau_c)}{\Gamma(1-\alpha,\:\lambda_2\tau_\mathrm{min})}.
\end{equation}

By knowing the probability, $p$, we can calculate how many interface defects are expected to be created before one has a lifetime of at least $\tau_c$
\begin{equation}
    \label{eq:ND}
    \langle N_\mathrm{D}\rangle = \frac{1}{p} = \frac{\Gamma(1-\alpha,\:\lambda_2\tau_\mathrm{min})}{\Gamma(1-\alpha,\:\lambda_2\tau_c)}.
\end{equation}

Since $\expND$ is independent of the expected time between the formation of interface defects, $\expdt$, and we know the supply rate of carbon atoms, $k$, we can calculate how many carbon atoms are expected to be added to the growing tube before an interface defect gets trapped
\begin{equation}
    \label{eq:NC}
    \langle N_\mathrm{C}\rangle = \langle N_\mathrm{D}\rangle\cdot\expdt\cdot k = \frac{k}{\lambda_1}\frac{\Gamma(1-\alpha,\:\lambda_2\tau_\mathrm{min})}{\Gamma\left(1-\alpha,\:\lambda_2\frac{2n_6}{k}\right)}.
\end{equation}

To calculate $\expNC$ using Eq.~\eqref{eq:NC}, we must first assume a value for $n_6$. If we consider the worst-case scenario in which an interface defect is formed such that only one 6-ring needs to be added to the interface of the growing tube to trap the defect, we get $n_6=1$. We can verify this assumption using Eq.~\eqref{eq:ND} with parameter values for the interface defect formation and lifetime distributions obtained from the growth of the $(6,5)$ SWCNT shown in Fig.~\ref{fig:growth}, $\lambda_1 = \pow{1.08}{9}$ \s{}, $\alpha = 1.20$, $\lambda_2 = \pow{1.04}{9}$ \s{}, $\tau_\mathrm{min} = \pow{1.10}{-12}$ s and $k=0.5$ \ns{}. This gives $\expND\approx 5991$ interface defects are expected to form before one gets trapped inside the tube wall, significantly more than the 804 interface defects seen during growth, Sim. 1 in Table~\ref{tab:defect_stat}. This confirms that $n_6=1$ is a reasonable assumption and we can now explore $\expND$ and $\expNC$ for different values of $k$.

For a carbon supply rate of $k=5$ \ns{}, ten times the value used to grow the $(6,5)$ SWCNT shown in Fig.~\ref{fig:growth}, we get $\expND\approx 19$ and $\expNC\approx 89$. Showing that at these high growth rates, interface defects are likely to live long enough to get trapped inside the tube wall, resulting in a defective CNT. This is consistent with our simulations at similar growth rates which yielded defective tubes, see Fig. S4. If, on the other hand, we use a carbon supply rate of $k=0.05$ \ns{}, one tenth of that used to grow the $(6,5)$ SWCNTs in Fig.~\ref{fig:growth} and similar to experimental growth rates, we obtain $\expND\approx \pow{1.425}{21}$ and $\expNC\approx \pow{6.599}{19}$. The key to growth of ultralong defect-free CNTs is thus to supply carbon atoms slow enough so that interface defects have time to heal and not get trapped inside the tube wall.

We can understand how reducing the growth rate enables growth of ultralong defect-free CNTs by looking at the CDF in Fig.~\ref{fig:growth}F. The slower a CNT grows, the longer an interface defect must live to get trapped and moving to the right in Fig.~\ref{fig:growth}F we see that for longer interface defect lifetimes the CDF approaches 1. Which corresponds to the survival function $S_\tau = 1 - F_\tau$ approaching 0 and thus it becomes extremely unlikely that an interface defect lives long enough to get trapped inside the tube wall.
\begin{figure}
    \centering
    \includegraphics{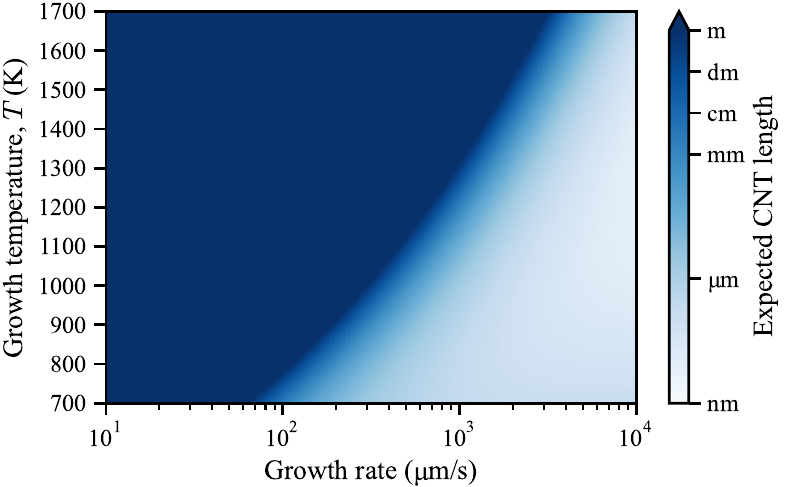}
    \caption{\label{fig:map} The expected length of defect-free CNTs for different combinations of growth rates and temperatures. Here Eq.~\eqref{eq:NC} was used to calculate the expected length and the result was converted to meters through multiplication with $\pow{8.35}{-12}$ m per C atoms. This value represents the length per C atom for a $(11,3)$ SWCNT with a diameter of 1 nm.}
\end{figure}

As seen in Table~\ref{tab:defect_stat} both the time between the formation of interface defects, $\delta t$, and their lifetimes, $\tau$ are affected by the growth temperature, $T$. Using simple analytical expressions, we can model the temperature behavior of $\lambda_1$, $\alpha$, $\lambda_2$ and $\tau_\mathrm{min}$ (see Supplementary material) and together with Eq.~\eqref{eq:NC} construct a map over the estimated CNT length for different combinations of growth temperatures and growth rates.

From this map, shown in Fig.~\ref{fig:map}, it is clear that low growth temperatures require a low growth rate, i.e., a low carbon supply rate in order to produce ultralong defect-free CNTs. Remarkably, however, Fig.~\ref{fig:map} also shows that increasing the growth temperature enables the growth of ultralong defect-free CNTs at significantly higher growth rates. Thus, if one can carefully tune the experimental conditions as to control the supply of carbon atoms (control the decomposition of the feedstock gas) and keep the catalyst stable (prevent Ostwald ripening, and melting) at high temperatures. One can achieve growth of ultralong defect-free CNTs at an order of magnitude higher growth rates compared to at low temperatures.

This result may seem counterintuitive given that high growth temperatures result in a decrease in $\expdt$ leading to the formation of more interface defects, as shown in Table~\ref{tab:defect_stat}. However, the decrease in the expected interface defect lifetime, $\exptau$, at high temperatures reduces the probability that these defects become trapped inside the tube wall during growth, counteracting the increased formation of interfacial defects at high temperatures. As seen in Eq.~\eqref{eq:NC}, the contribution of $\expdt$ to the expected CNT length is proportional to $\frac{k}{\lambda_1}$ and decreases with increasing $T$, due to the increase in $\lambda_1$ with $T$. On the other hand, the contribution of $\exptau$ to the expected CNT length is proportional to the ratio of two upper incomplete gamma functions, $\frac{\Gamma(1-\alpha,\:\lambda_2\tau_\mathrm{min})}{\Gamma\left(1-\alpha,\:\lambda_2\frac{2n_6}{k}\right)}$. For which the numerator remains largely unaffected by changes in $T$ as the product $\lambda_2\tau_\mathrm{min}$ stays small even as $\lambda_2$ increases with $T$. The denominator however rapidly approaches zero as the term $\lambda_2\frac{2n_6}{k}$ increases with $T$, causing the ratio of the upper incomplete gamma functions to overwhelm the decreasing $\frac{k}{\lambda_1}$ term.

Our results show that there is no theoretical limit to how long defect-free CNTs can be grown if the growth rate matches the growth temperature, and the experimental conditions are stable. Moreover, higher growth temperatures enable faster growth of ultralong defect-free CNTs if the carbon supply can be carefully controlled.

\subsection{Dynamics of growing nanotube interfaces}
\begin{figure*}
    \centering
    \includegraphics[width=\textwidth]{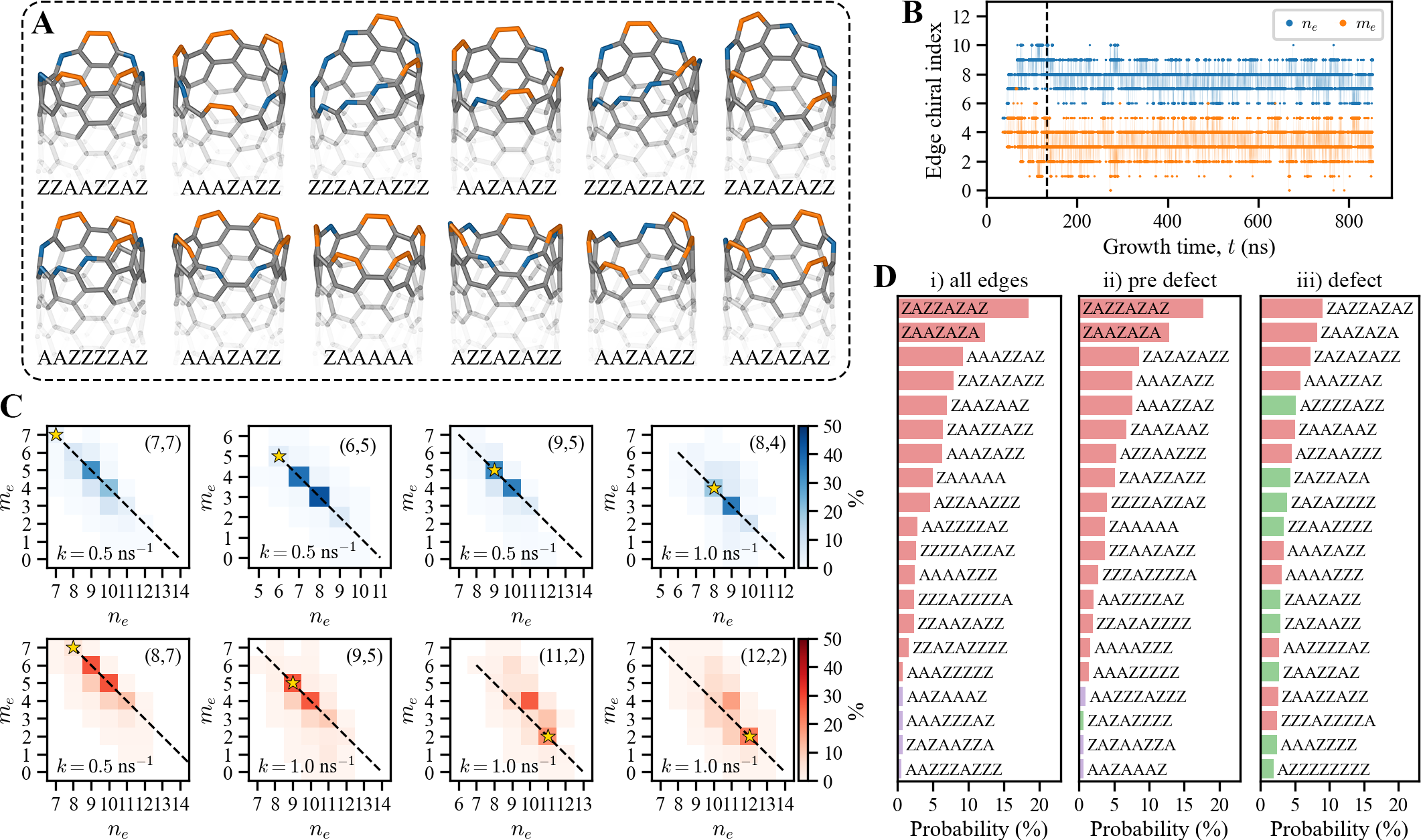}
    \caption{\label{fig:edge} Edge configurations observed during growth of SWCNTs on a \Fe{} catalyst. \textbf{A} selection of unique edges present during the growth of the $(6,5)$ SWCNT shown in Fig.~\ref{fig:growth}. Here zigzag sites are denoted by Z and colored blue in the structure while armchair pairs are denoted by A and colored orange. \textbf{B} time series of the edge chiral index, $(n_e,m_e)$, during growth of the $(6,5)$ SWCNT. The dashed vertical line marks the time at which SWCNT-cap is fully formed, $t=132.41$ ns. \textbf{C} 2D histograms showing the distribution of edge chiral indices for different tubes grown at 1300 K (blue) and 1500 K (red). Here the chirality $(n,m)$ of the final grown tube is shown in the upper right corner of the 2D histograms and marked by the gold star. The dashed line shows where the length of the edge, $n_e+m_e$, matches the length of a perpendicular cut edge of the final grown tube, $n+m$. For each SWCNT it's growth rate is denoted by $k$ in the histograms. \textbf{D} bar graph of the 20 most observed edge configurations during i) the entire growth simulation of the $(6,5)$ SWCNT, ii) just before the formation of an interface defect and iii) after the formation of an interface defect. Here the color of the bars represents the length of the edge where green: $n_e+m_e = 10$, red: $n_e+m_e = 11$ and purple: $n_e+m_e = 12$.}
\end{figure*}
\begin{figure*}
    \centering
    \includegraphics[width=\textwidth]{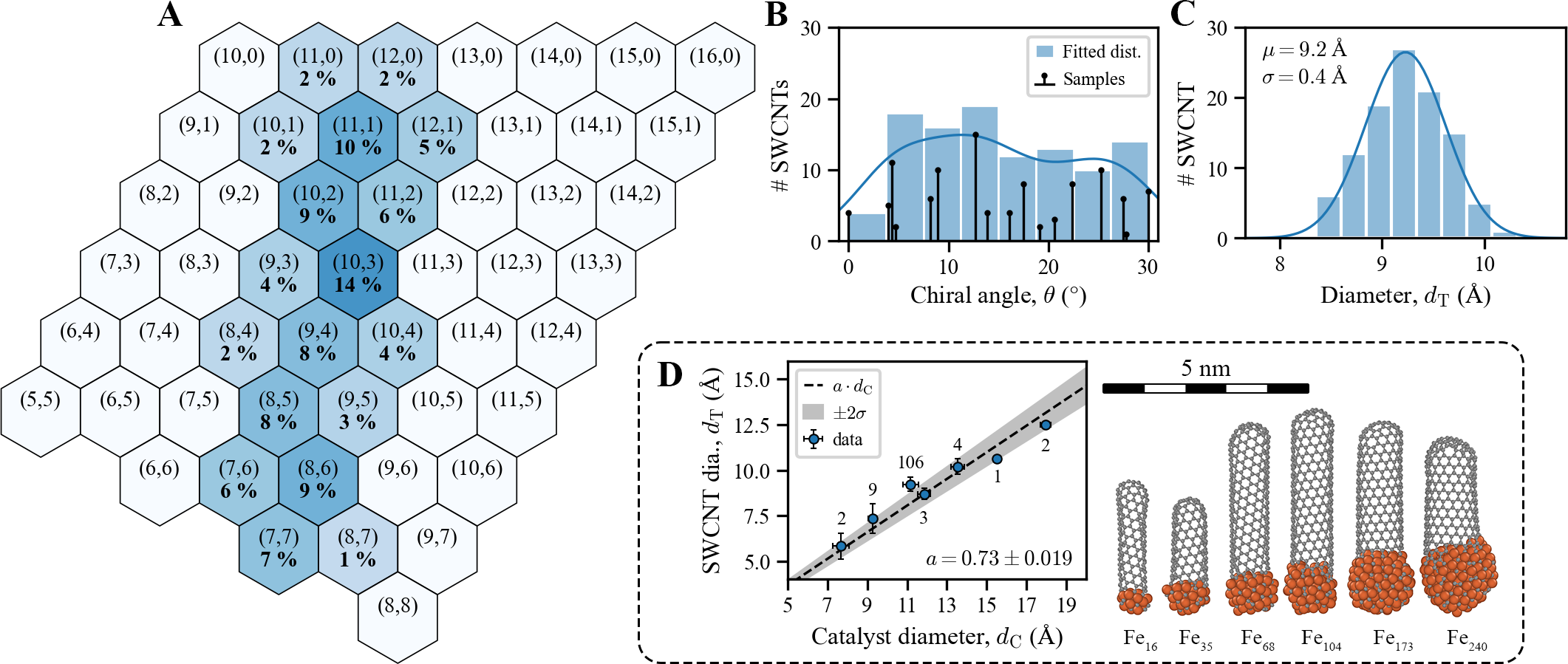}
    \caption{\label{fig:chiral} Chirality distribution and diameter dependent growth. \textbf{A} chirality map of 106 individual SWCNTs grown using a \Fe{} catalyst at $T=1300$ K with a carbon supply rate of $k=0.5$ \ns{}. The percentage shows the abundance of each chirality $(n,m)$ observed after growth. \textbf{B}, \textbf{C} the chiral angle and SWCNT diameter distributions, respectively, for the 106 SWCNTs in \textbf{A}. Panel \textbf{D} show SWCNTs grown on different sized Fe catalysts with diameters, $d_C$, here the resulting tube diameters, $d_T$, follows a linear relationship $d_T = a\cdot d_C$ where $a = 0.73\pm 0.019$. The gray bands indicate the $2\sigma$ confidence interval of the fit and the number of SWCNTs grown for each catalyst size is marked above the data points. Visualized atomic configurations show examples of SWCNTs grown on the different sized catalysts.}
\end{figure*}

The evolution of the SWCNT-edge during growth has long been a topic of interest, with early work suggesting that the edge evolves in a screw dislocation fashion\cite{ding2009dislocation}. Later research recognized the importance of configurational entropy\cite{2018entropy-driven} and the impact of curvature and armchair/zigzag junctions\cite{liu2010graphene, hedman2020analytical}. More recent advances have extended the concept of configurational entropy to include edges of any cut where the number of edge atoms remains constant\cite{forster2023swinging}. This model allows for more realistic growth modes, as the edge is not limited to evolve in a screw dislocation fashion.

In our MD simulations, we can follow the evolution of the SWCNT-edge during growth by tracking it's configuration, see Supplementary material for more details. The entire evolution of the edge, after the formation of the SWCNT-cap, for the growth of our showcase $(6,5)$ SWCNT can be viewed in \href{https://youtu.be/JFKhklSHgA4}{Supporting video S4}. From which Fig.~\ref{fig:edge}A shows a selection of unique edge configurations present during growth, highlighting the highly dynamic nature of the edge. Our results show that the edge exhibits multiple configurations during growth with varying numbers of armchair pairs, $N_A$, and zigzag sites, $N_Z$, and does not evolve in a continuous screw dislocation fashion.

To better understand the evolution of the edge during growth, we derive an edge chiral index, $(n_e,m_e)$, where $n_e=N_A+N_Z$ and $m_e=N_A$. This edge chiral index allows for easy identification of edges with the same number of armchair pairs and zigzag sites as a perpendicular cut edge on a tube with chirality $(n,m)$, since for these, $n_e=n$ and $m_e=m$. As seen in Fig.~\ref{fig:edge}B, $(n_e,m_e)$ fluctuates significantly during growth, even after the SWCNT-cap is fully formed. Where $m_e$ varies from zigzag edge $(m_e=0)$ with the aid of a 5-ring interface defect to near zigzag $(m_e=1)$ and near armchair $(n_e-m_e=1)$.

The computational efficiency of DeepCNT-22 allows us to run multiple growth simulations to produce several defect-free SWCNTs, as shown in Fig. S3. These tubes are long enough to accurately determine the distribution of the edge chiral index during growth, as shown in Fig.~\ref{fig:edge}C. Interestingly, we find that the most dominant edge chiral index does not necessarily match the chirality of the grown tube. Instead, the SWCNT-edge seems to maximize the configurational entropy, in agreement with models proposed by Bichara et al.\cite{2018entropy-driven, forster2023swinging}. This results in the most common edge during growth being chiral, regardless of the chirality of the tube, as is evident by comparing the $(n_e,m_e)$ distributions for the $(7,7)$ and $(9,5)$ SWCNTs in Fig.~\ref{fig:edge}C. Our results also show that the distribution of edge chiral indices is largely unaffected by the growth temperature, as seen by comparing the distributions for $(9,5)$ SWCNTs at 1300 K (blue) and 1500 K (red) in Fig.~\ref{fig:edge}C.

The edge chiral index is not a unique identification of a SWCNT-edge since there are multiple ways to arrange $N_A$ armchair pairs and $N_Z$ zigzag sites, which is what gives rise to the configurational entropy of the edge. For the $(6,5)$ SWCNT, the most dominant edge with a probability of 43.3\% is $(n_e,m_e) = (8,3)$, closely followed by $(n_e,m_e) = (7,4)$ with a probability of 37.3\%. These two edge chiral indices account for 80.6\% of all edge chiral indices observed during growth thus, as discussed above, the SWCNT-edge seems to maximize the configurational entropy. However, to confirm this, it is necessary to determine whether there is a preferred edge configuration or set of configurations observed during growth. Thus, we count the occurrence of each unique edge configuration, taking into account the cyclic nature of the edge i.e., that AAAAZZZ is equivalent to ZZAAAAZ for example. As shown in Fig.~\ref{fig:edge}D i), the most frequently observed edge configuration during growth is ZAZZAZAZ with a probability of 18.5\%, closely followed by ZAAZAZA (12.4\%), AAAZZAZ (9.18\%), ZAZAZAZZ (7.89\%), and so on. Which proves that there is no preferred edge configuration or set of configurations during growth and confirms that the edge maximizes the configurational entropy. Further evidence of this can be found in Fig. S5 for the other tubes grown. Our findings thus confirm the importance of the configurational entropy of the SWCNT-edge during growth.

To relate these findings to the discussion of interface defects in the previous section, we use the edge configuration to determine if any particular edge leads to the formation of interface defects. By analyzing the configurations of the edge just before the formation of an interface defect, Fig.~\ref{fig:edge}D ii), and comparing with those of all the edges seen during growth, Fig.~\ref{fig:edge}D i), it is clear that no particular configuration stand out. This further supports the conclusion that the formation of interface defects is purely stochastic and does not depend on the configuration of the edge. After the formation of an interface defect the configurations of the edge, Fig.~\ref{fig:edge}D iii), are significantly different and shows several edges of shorter length (green bars) due to the formation of a 5-ring at the edge.

\subsection{Chirality distribution and diameter dependent growth}

Having shown that DeepCNT-22 can be used to grow long defect-free SWCNTs, we demonstrate the strength of this MLFF by studying the chirality distribution of SWCNTs grown on \Fe{} catalysts as well as how the diameter of grown tubes depends on the diameter of the catalyst.

Experimental works have successfully grown SWCNTs with super narrow chirality distributions using CoW catalysts and ethanol feedstock\cite{yang2014chirality-specific, yang2015growing, yang2016water-assisted, yang2022growth}. However, simpler growth experiments using monometallic catalysts such as Fe or Ni and hydrocarbon feedstock gas have shown broad chirality distributions\cite{chiang2009linking, he2012diameter}. These experimental growth conditions are comparable to our simulations, and thus we expect to be able to reproduce similarly broad chirality distributions. To verify this, we performed 280 growth simulations each at identical growth condition; \Fe{} catalyst at a temperature of 1300 K with 200 carbon atoms supplied at a rate of $k=0.5$ \ns{}.

Our simulations resulted in 106 SWCNTs with well-defined chirality, as shown in Fig. S6, corresponding to a yield of 37.9\%. The 174 tubes with undefinable chirality can be seen in Fig. S7 where growth failed in various ways from partially encapsulated catalysts to cone-shaped caps and trapped defects. Fig.~\ref{fig:chiral}A presents the chirality map of the 106 tubes with well-defined chirality which shows a broad spread in chirality, as confirmed by the chiral angle distribution in Fig.~\ref{fig:chiral}B. Both the map and the chiral angle distribution reveal an underrepresentation of zigzag tubes, which may be due to either a lower frequency of nucleation for these tubes or a lower probability of survival during the growth process. The diameter distribution of the grown tubes, shown in Fig.~\ref{fig:chiral}C, has a mean of 9.2 Å and a standard deviation of 0.4 Å, which is in good agreement with the experimental results of Ago et al.\cite{ago2005cvd}, who measured a mean diameter and standard deviation of 9.3 and 0.6 Å, respectively.

By growing SWCNTs on catalysts of different sizes we can determine the ratio between the diameter of the grown tubes, $d_T$, and that of the catalysts, $d_C$. Here the growth temperature is the same $T=1300$ K for each catalyst size, but the total number of carbon atoms added and the supply rate, $k$, was tuned for each catalyst size to achieve growth. Fig.~\ref{fig:chiral}D shows the mean SWCNT diameter as function of the mean catalyst diameter where a linear relationship can be found $d_T = a\cdot d_C$ where $a = 0.73\pm 0.019$ in agreement with previous experimental\cite{diaz2019single-walled, yang2022growth} and theoretical\cite{ding2004molecular, xu2021catalyst} studies.

\section{Conclusions}

This study presents a novel machine learning force field (MLFF), DeepCNT-22, for molecular dynamics (MD) simulations of single-walled carbon nanotube (SWCNT) growth on iron catalysts. DeepCNT-22 can simulate the complete growth process with high accuracy at close to microsecond timescales, enabling growth of long defect-free SWCNTs for the first time without the use of steering or other biases.

Our simulations reveal the five stages of SWCNT growth along with new atomic-level insights, including the formation and healing of defects and the evolution of the SWCNT-edge. By analyzing our growth simulations, we show that 5- and 7-ring defects form stochastically at the SWCNT-edge, and we model the formation and lifetime of these interface defects statistically. From these new insights, combined with the growth rate and temperature, we propose a model to predict how long defect-free CNTs can be expected to grow. Showing that if the growth rate is appropriately chosen based on the growth temperature, there is in practice no theoretical limit on how long defect-free CNTs can grow. Additionally, our model suggest that high growth temperatures can enable growth of ultralong defect-free CNTs at an order of magnitude higher rates, if the supply of carbon atoms can be controlled. Furthermore, we analyze the evolution of the SWCNT-edge during growth and show that it is highly dynamic in nature, displaying a variety of different configurations of armchair pairs and zigzag sites. Remarkably we found that the chirality of the edge does not necessarily match the chirality of the grown tube, instead maximizing configurational entropy. The power of DeepCNT-22 is demonstrated by the growth of hundreds of SWCNTs whose chirality emerges naturally from our simulations and shows a broad chirality distribution consistent with growth experiments on iron catalysts. In conclusion, this study highlights the effectiveness of MLFFs in uncovering the underlying mechanisms of SWCNT growth and offers a path towards realizing the full potential of SWCNTs in future technologies.

\section{Methods}

The dataset used for training of the DeepCNT-22 MLFF and the entire MD trajectory for the growth simulation of the $(6,5)$ SWCNT will be made available online at the time of journal publication. Supporting videos \href{https://youtu.be/K90Ca6uDNEQ}{S1}, \href{https://youtu.be/x8Z5Go5iW58}{S2}, \href{https://youtu.be/e1Yx14PQjkg}{S3} and \href{https://youtu.be/JFKhklSHgA4}{S4} were created from the MD trajectory of the grown $(6,5)$ SWCNT. After the simulation, the SWCNT structure was aligned such that it's axis is parallel to the z-axis and the Supporting videos were then rendered from the aligned MD trajectory using OVITO software\cite{stukowski2009visualization}. The \texttt{Smooth trajectory} modifier with a window size of 5 was used to reduce thermal vibrations, making it easier to track the evolution of the structure during growth. The evolution of the SWCNT-edge during growth, as shown in \href{https://youtu.be/JFKhklSHgA4}{Supporting video S4}, was visualized by first removing all iron atoms, and then iteratively removing carbon atoms with a coordination number of less than 2 until only atoms with a coordination number of 2 or higher remained. The edge atoms were then colored as follows, blue: zigzag sites, orange-orange: armchair pairs, and orange-green-orange: directly nucleated 6-rings.

\subsection{Density functional tight binding}

The initial dataset for development of DeepCNT-22 comprised of structures obtained from DFTB MD simulations of SWCNT nucleation from atomic carbon precursors on Fe nanoparticle catalysts. DFTB is an extended two-center Hückel approximation to DFT utilizing a minimal Slater-type all valence basis set, allowing for dynamic simulation orders of magnitude faster than DFT while still including electronic effects unlike classical force field-based methods. MD simulations utilized self-consistent charge DFTB (SCC-DFTB)\cite{elstner1998self-consistent-charge} to calculate quantum chemical potential energy and energy gradients on the fly for each MD iteration. The \texttt{trans3d-0-1} parameter set was used\cite{zheng2007parameter}, with all simulations employed within the DFTB+ software package\cite{hourahine2020dftb+} version 21.1. Newton's equations of motion were integrated with the velocity-Verlet algorithm\cite{swope1982computer}, with a time step of 1.0 fs and a finite electronic temperature of 10 000 K\cite{weinert1992fractional, wentzcovitch1992energy, wagner1998errors}. A canonical NVT ensemble was enforced at a temperature of 1500 K using a Nosé-Hoover chain thermostat\cite{nose1984unified, hoover1985canonical, martyna1992nose-hoover} of length 3.

Structures were obtained from MD simulations initially containing a Fe\textsubscript{13}, Fe\textsubscript{38} or Fe\textsubscript{55} nanoparticle in a periodic cell in the absence of C atoms, or containing 20, 30 or 40 C atoms for Fe\textsubscript{13} at a minimum distance of 5 Å. Structures extracted from these simulations included Fe nanoparticles with surface-adsorbed carbon monomers and dimers, carbon chains and junctions, ring networks often comprising defects and SWCNT-cap and tube-like structures, in agreement with prior DFTB growth simulations\cite{ohta2009quantum, mclean2022mechanism}. DFTB MD simulations were also performed to anneal high energy structures obtained by early version of our MLFF with resulting structures added to the dataset. To determine which structures from the DFTB MD simulations to label with DFT and add to the training data, farthest point sampling was performed on the DFTB calculated potential energy.

\subsection{Density function theory}

DFT calculations were carried out using the Vienna Ab initio Simulation Package (VASP)\cite{kresse1993ab, kresse1996efficiency, kresse1996efficient} version 6.3.0. A plane wave basis set was employed, with the projector-augmented wave method\cite{blochl1994projector, kresse1999ultrasoft} used in conjunction with standard pseudopotentials (\texttt{Fe 06Sep2000} and \texttt{C 08Apr2002}). The \texttt{optB86b-vdW} van der Waals density functional\cite{dion2004van, klimes2011van} was chosen to account for dispersion interactions. High precision (\texttt{PREC = Accurate}) was used throughout the calculations, with a plane wave cutoff energy of 600 eV (\texttt{ENCUT = 600}) and no symmetry constraints applied (\texttt{ISYM = 0}). To ensure the accuracy of the results, the electronic self-consistent loop was converged to a tolerance of $10^{-6}$ eV (\texttt{EDIFF = 1.0E-6}). Gaussian smearing (\texttt{ISMEAR = 0}) was used with a smearing width of 0.05 eV (\texttt{SIGMA = 0.05}) to aid in the convergence of the calculations. Spin-polarized calculations were performed (\texttt{ISPIN = 2}), with a high initial magnetic moment, 3 e$^-$, assigned to each Fe atom. For all periodic structures, a $\Gamma$-centered k-point mesh with a density of 0.25 Å$^{-1}$ (\texttt{KSPACING = 0.25}) was used, while for the non-periodic structures, only the $\Gamma$-point was used with at least 10 Å vacuum spacing between the periodic images. Since DFT calculations were used to label the training data, only single point calculations were performed.

\subsection{Machine learning force field}
DeepCNT-22 is based on the Deep Potential Smooth Edition architecture\cite{zhang2018end-to-end} and was developed using DeePMD-kit\cite{wang2018deepmd-kit:} version 2.1.1. This MLFF is of the Behler-Parrinello type\cite{behler2007generalized}, in which the energy of each atom in a structure is predicted using a neural network, and the atomic energies are then summed to give the total energy of the structure. Here, we used a type map of \texttt{[Fe, C]} and employed the type embedding approach, which improves performance and accuracy by allowing the use of a single descriptor embedding net and fitting net shared by both atom types. For more information on the Deep Potential Smooth Edition architecture and the type embedding approach, refer to the DeePMD-kit documentation\cite{wangdeepmd-kits}.

For the type embedding approach we used an embedding net with 2 hidden layers containing 8 neurons each. The descriptor embedding net was of type \texttt{se\_e2\_a} and comprised of 3 hidden layers with 16, 32, and 64 neurons, as well as 8 axis neurons. We applied a cutoff of 5.0 Å to define the local environment of each atom, with a smooth cutoff of 0.5 Å and used a fitting net consisting of 3 hidden layers with 256 neurons each. The activation function for each hidden layer was the GELU function\cite{hendrycks2016gaussian}, and no timestep was used in the ResNet architecture\cite{he2015deep}. For training we applied the following loss function $\mathcal{L} = \frac{p_\epsilon}{N}\Delta E^2 + \frac{p_f}{3N}\left|\Delta\boldsymbol{F}\right|^2$, where $N$ is the number of atoms, $E$ is the energy, and $\boldsymbol{F}$ are the forces of each structure. Here, the energy and force error weights, $p_\epsilon$ and $p_f$, were set to 0.1 and 1.0, respectively, and these values were kept fixed during training. Training was performed for 300 000 batches, using a batch size of 5 structures and the Adam optimizer\cite{kingma2014adam:} with an initial learning rate of $10^{-3}$ which was decayed exponentially to $10^{-5}$ at the end of the training.

\subsection{Molecular dynamics}

MD simulations of SWCNT growth were performed using the Large-scale Atomic/Molecular Massively Parallel Simulator (LAMMPS)\cite{thompson2021lammps} version 29 Sep 2021 - Update 3 with the \texttt{deepmd} pair style and the DeepCNT-22 MLFF. The \texttt{nsq} algorithm, with a cutoff distance of 5.0 Å, was employed to construct the neighbor list, as it is slightly faster for smaller systems. An additional 2.0 Å skin distance was utilized, and the neighbor list was rebuilt only if at least one atom had moved more than half the skin distance. Simulations were performed in the NVT ensemble using a Nosé-Hoover chain thermostat\cite{nose1984unified, hoover1985canonical, martyna1992nose-hoover} of length 3, with the temperature damping parameter set to 0.1 ps. The equations of motion were integrated using a timestep of 2.0 fs, which maximizes performance while maintaining the stability of the simulation. Initial velocities for the atoms were drawn from a Gaussian distribution and the resulting ensemble of velocities had their linear and angular momenta zeroed, the velocities were then scaled to correspond to the growth temperature $T$. The mass for the Fe and C atoms was set to 55.847 u and 12.011 u, respectively.

Carbon atoms were deposited one by one at a rate of $k$ \ns{}, with the deposit region being a sphere, located at the center of the simulation box, of diameter $d_C/2$, where $d_C$ is the diameter of the Fe catalyst. To ensure that carbon atoms were always deposited inside the iron catalyst, the system was recentered at every timestep such that the center of the catalyst is always at the center of the simulation box. The number of degrees of freedom contributing to the system temperature was dynamically updated to account for the deposited carbon atoms. Simulation data was written to files every 2 ps for later analysis, including the system temperature, potential energy, number of atoms with carbon-carbon coordination numbers 0, 1, 2, and 3, and the total number of carbon atoms added, as well as atomic coordinates, energies, and carbon-carbon coordination numbers. For details on how this data was analyzed, see the Supplementary material.

\section{Acknowledgments}

The authors would like to acknowledge the computational resources provided by the Swedish National Infrastructure for Computing via the SNIC 2022/3-29 and SNIC 2022/5-110 projects, partially funded by the Swedish Research Council through grant agreement no. 2018-05973. As well as the computational resources provided by the Institute for Basic Science (Korea) at the HPC clusters Cimulator (CMCM, Ulsan) and Olaf (IBS-HQ, Daejeon).

\bibliography{main}

\end{document}